\documentstyle[twocolumn,floats,prd,aps]{revtex}

\begin{document}
\draft
\preprint{nucl-ex/9811001}
\title{Electroproduction of the
$S_{\bbox{11}}(1535)$ Resonance at High Momentum Transfer}
\author{C.~S.~Armstrong$^{2}$\cite{csa_byline}, P.~Stoler$^{7}$,
G.~S.~Adams$^{7}$, A.~Ahmidouch$^{3,4}$, K.~Assamagan$^{3}$,
S.~Avery$^{3}$, O.~K.~Baker$^{3,8}$, P.~Bosted$^{1}$,
V.~Burkert$^{8}$, R.~Carlini$^{8}$, J.~Dunne$^{8}$,
T.~Eden$^{3}$, R.~Ent$^{8}$, V.~V.~Frolov$^{7,}$\cite{vf_byline},
D.~Gaskell$^{3}$, P.~Gu\`{e}ye$^{3}$, W.~Hinton$^{3}$,
C.~Keppel$^{3,8}$, W.~Kim$^{5}$, M.~Klusman$^{7}$,
D.~Koltenuk$^{9}$, D.~Mack$^{8}$, R.~Madey$^{3,4}$,
D.~Meekins$^{2}$, R.~Minehart$^{10}$, J.~Mitchell$^{8}$,
H.~Mkrtchyan$^{11}$, J.~Napolitano$^{7}$, G.~Niculescu$^{3}$,
I.~Niculescu$^{3}$, M.~Nozar$^{7}$,
J.~W.~Price$^{7,}$\cite{jp_byline}, V.~Tadevosyan$^{11}$,
L.~Tang$^{3,8}$, M.~Witkowski$^{7}$, S.~Wood$^{8}$\\[5pt]}

\address{
\emph{
$^{1}$\emph{Physics Department, American University, Washington
D.C. 20016, USA} \\
$^{2}$\emph{Department of Physics, College of William \& Mary,
Williamsburg, VA 23187, USA} \\
$^{3}$\emph{Physics Department, Hampton University, Hampton, VA
23668, USA} \\
$^{4}$\emph{Physics Department, Kent State University, Kent, OH
44242, USA} \\
$^{5}$\emph{Physics Department, Kyungpook National University,
Taegu, South Korea} \\
$^{7}$\emph{Physics Department, Rensselaer Polytechnic Institute,
Troy, NY 12180, USA} \\
$^{8}$\emph{Thomas Jefferson National Accelerator Facility,
Newport News, VA 23606, USA} \\
$^{9}$\emph{Physics Department, University of Pennsylvania,
Philadelphia, PA 19104, USA} \\
$^{10}$\emph{Physics Department, University of Virginia,
Charlottesville, VA 22903, USA} \\
$^{11}$\emph{Yerevan Physics Institute, Yerevan, Armenia} \\[5pt]
}}
\date{\today}
\maketitle
\begin{abstract}
The differential cross section for the process $p(e,e'p)\eta$ has
been measured at $Q^2 = 2.4$ and 3.6~(GeV$/c)^2$ at
center-of-mass energies encompassing the $S_{11}(1535)$
resonance. The latter point is the highest-$Q^2$ exclusive
measurement of this process to date.  The resonance width and the
helicity-1/2 transition amplitude are extracted from the data,
and evidence for the possible onset of scaling in this reaction
is shown. A lower bound of $\approx 0.45$ is placed on the
$S_{11}(1535) \rightarrow p\,\eta$ branching fraction.

\end{abstract}
\narrowtext

\section{Introduction}
\label{sec:intro}

Baryon electroproduction allows the measurement of transition
form factors, which test models of hadronic structure in ways
that static baryon properties alone cannot. Recently much effort
has gone into attempts to reproduce observed transition form
factors over a large range of four-momentum transfer. At low
four-momentum transfer, or $Q^2$, the focus has been on
incorporating relativistic effects into the constituent quark
model (CQM)~\cite{Isg77}, using
light-front~\cite{Cap95,Cap98,Sta95} and other~\cite{Li90,War90}
approaches. At higher $Q^2$, perturbative QCD (pQCD) sum rule
calculations~\cite{Car88} and valence pQCD~\cite{valpqcd} have
been employed. The applicable range in $Q^2$ for these various
approaches is not clear.

Among the most interesting of baryon case studies is the
$S_{11}(1535)$ resonance, which is one of the most strongly
excited states over all $Q^2$, and which is easily isolated
because it is the only resonance that has a large branching
fraction to the $\eta$. The reproduction of the $S_{11}(1535)$
form factor has become a goal of many models, but the effort has
been hampered by a lack of precise electroproduction data. In
addition, the uncertainty in the $S_{11}(1535)$ transition
amplitude is limited by knowledge of the full width and branching
fraction to the $\eta$.  We report here on a measurement of the
reaction $e+p \rightarrow e'+S_{11}(1535) \rightarrow e'+p+\eta$
and an extraction of the helicity-conserving transition amplitude
$A^p_{1/2}$ at $Q^2 = 2.4$ and 3.6~(GeV$/c)^2$. We also use a
recent analysis of inclusive $(e,e')$ data to put a lower bound
on the $S_{11}(1535) \rightarrow p\,\eta$ branching fraction.

\section{The Experiment}
\label{sec:exp}

The experiment was performed in Hall C of the Thomas Jefferson
National Accelerator Facility (Jefferson Lab), shown in
Figure~\ref{fig:hallcfloorplan}. The Short Orbit Spectrometer
(SOS), which is a resistive $QD\bar{D}$ device, was used to
detect electrons. The High Momentum Spectrometer (HMS), which is
a superconducting $QQQD$ spectrometer, was used to detect
protons.  Figure~\ref{fig:hdetstack} shows the HMS detectors,
which include drift chambers (DC1 and DC2) for determining track
information, scintillator arrays (S1X/Y and S2X/Y) for triggering
and time-of-flight measurement, and a threshold gas \v{C}erenkov
and electromagnetic calorimeter for particle identification
(PID). The SOS detectors are configured similarly.

The incident electrons had energies $E=3.245$ and 4.045~GeV for
the $Q^2 = 2.4$ and 3.6~(GeV$/c)^2$ points, respectively.  At
each of the two $Q^2$ points, the electron spectrometer was fixed
in angle and momentum, thus defining a central three-momentum
transfer $\mathbf{q}$ and direction of a boosted decay cone of
protons.  The proton spectrometer was stepped in angle and in
momentum to capture as much of this decay cone as possible. Data
were obtained at 33 (21) kinematic settings at the low (high)
$Q^2$ point.

Target protons were provided in the form of liquid hydrogen at
19~K flowing through a target of length 4.36~cm. The relative
current of the electron beam was measured by two resonant-cavity
current monitors, which were calibrated periodically using the
absolute beam current measurement of a parametric current
transformer. The combined measurement had an absolute accuracy of
\mbox{$\sigma = 1.5$\%}.

Electrons were identified in the SOS using the \v{C}erenkov
detector and lead-glass calorimeter (see
Figure~\ref{fig:sospid2d}). In the HMS, protons were separated
from pions using the time of flight measured between two pairs of
scintillator arrays (see Figure~\ref{fig:cointime2d}).  In both
spectrometers, tracking information was obtained from the drift
chambers. Details of the experiment and analysis are given in
Ref.~\cite{Arm98}, and information on a simultaneous measurement
of the $\Delta(1232)$ can be found in Ref.~\cite{Fro98}.

\section{Data Analysis}
\label{sec:datanal}

The data were corrected for trigger and PID inefficiencies
(\mbox{$< 1$\%}), track reconstruction inefficiencies
(\mbox{$\approx 5$\%}), computer and electronic dead times
(\mbox{$< 5$\%}), current-dependent target density changes
(\mbox{$\approx 3$\%}), and protons undetected due to
interactions in the detector stack (\mbox{$\approx 3$\%}). The
data were binned in $W$, $\cos{\theta^*_{\eta}}\,$,
$\phi^*_{\eta}\,$, and $M_{x}^2$ (with 6, 10, 6, and 20 bins,
respectively). Here $W$ is the invariant mass; $\theta^*_{\eta}$
is the polar angle between the direction of the $\eta$ and the
three-momentum transfer $\mathbf{q}$ in the center-of-mass
(c.m.)\ of the resonance; $\phi^*_{\eta}$ is the azimuthal angle
of the $\eta$ with respect to the electron scattering plane; and
$M_{x}^2$ is the square of the missing mass for $p(e,e'p)X$. The
$\eta$ mesons were identified in the final state using
$M_{x}^2$. Figure~\ref{fig:fridgeplot} shows the missing mass
distribution for a typical kinematic setting.

Modest backgrounds in $M_{x}^2$ due to accidentals
(\mbox{$\approx 2$\%}, shown in Figure~\ref{fig:cointime2d}) and
protons penetrating the HMS collimator and magnet apertures
(\mbox{$\approx 4$\%}) were measured and subtracted from the
data. The remaining continuum background in missing mass was due
to multi-pion ($n\pi$) production (ranging from $30$\% to $50$\%
of the resonance data) and a small (\mbox{$< 2$\%}) contribution
from target-window interactions. Two independent techniques were
used to subtract these remaining background events. The first
technique fitted a polynomial plus peak in $M_{x}^2$ to the data
in each ($\cos{\theta^*_{\eta}}$, $\phi^*_{\eta}$) bin
(integrated over the $W$ acceptance for that kinematic setting),
and then subtracted the background contribution from each bin.
The second technique scaled a Monte Carlo-generated $n\pi$
background to match the data above and below the missing-mass
peak and then subtracted this background from each ($W$,
$\cos{\theta^*_{\eta}}$, $\phi^*_{\eta}$) bin.

Three different models were used to simulate the $n\pi$
background in the Monte Carlo: $e\,p \rightarrow
e'\,p\,\pi^+\,\pi^-$, $e\,p \rightarrow e'\,\Delta^{++}\,\pi^-
\rightarrow e'\,p\,\pi^+\,\pi^-$, and a crude approximation of
three-body phase space. The Monte Carlo simulation was also used
to simulate multiple scattering and ionization energy loss, and
to correct for experimental acceptance and the effect of
radiative processes. Once the $n\pi$ background was subtracted
from both experimental and simulated spectra, the experimental
yields were corrected to account for finite $Q^2$ acceptance. The
differential cross section was then given by the ratio of
experimental to simulated yield in each ($W$,
$\cos{\theta^*_{\eta}}$, $\phi^*_{\eta}$) bin, normalized by the
simulation resonance cross section for that bin.

The cross sections obtained using the different $n\pi$ models and
the two background subtraction techniques all agreed within
$2$\%; both the following figures and our final results were
obtained using the first subtraction technique together with a
background generated by combining two of the $n\pi$ models.
Figure~\ref{fig:indivfit} shows data and fits for several typical
($\cos{\theta_{\eta}^*}\,$, $\phi^*_{\eta}$) bins of one
kinematic setting. Figure~\ref{fig:bgfit1} shows the result of
fits for several kinematic settings, where for each setting we
have integrated both the data and their respective fits over the
sixty individual ($\cos{\theta_{\eta}^*}\,$, $\phi^*_{\eta}$)
bins.

Using similar techniques we verified the well-known
$^1\mathrm{H}(e,e'p)$ cross section~\cite{elast} to within $2$\%.

\section{Results}
\label{sec:results}

The five-fold differential cross section for the $e\,p
\rightarrow e'p\,\eta$ process may be expressed as the product of
the transverse virtual photon flux $\Gamma_{_T}$ (Hand
convention~\cite{Han63}) and the c.m.\ cross section for the
electroproduction of the $p\,\eta$ pair:
\begin{equation}
\label{eq:d5sigma}
\frac{d \sigma}{d\Omega_e dE'_e d\Omega^*_{\eta}} =
\Gamma_{_T}\ \frac{d \sigma}{d\Omega^*_{\eta}}\,(\gamma_v\,p
\rightarrow p\,\eta)\ .
\end{equation}
Previous data indicate that the c.m.\ $\gamma_v\,p \rightarrow
p\,\eta$ cross section is dominated by $S$-waves arising from the
$S_{11}(1535)$~\cite{Kru95,Bra84}.  This dominance was confirmed
by the present data, which showed that terms other than $S$-wave
were less than $7$\% and consistent with zero within the
statistical uncertainty of the data.  Angular distributions for
the $Q^2 = 3.6$~(GeV$/c)^2$ data are shown in
Figure~\ref{fig:angdist5pass}.

From $S$-wave fits to the angular distributions, the total cross
section was calculated (at each $Q^2$ point) as a function of
$W$:
\begin{equation}
\sigma_{\mathrm{tot}}(W) = 4 \pi\,
\frac{d \sigma}{d\Omega^*_{\eta}}(\gamma_vp\rightarrow
p\eta)\ .
\end{equation}
This cross section, which consists of resonant and nonresonant
parts, was fitted with a relativistic Breit-Wigner plus
nonresonant background curve,
\begin{eqnarray}
\label{eq:breitwig}
\sigma_{\mathrm{tot}}(W)\ & = &\
\sigma_{\mathrm{res}}(W) \ +\ \sigma_{\mathrm{nr}}(W)
\nonumber\\[0.1cm] \ & = & \ A_{\mathrm{res}}^2 \ 
\frac{|{\mathbf p}^*_{\eta}|\,W}{m_p\,K} \ 
\frac{W_R^2\,\Gamma_R^2}{(W^2-W_R^2)^2 + W_R^2\, \Gamma^2(W)}
\nonumber\\[0.1cm] & & +\ B_{\mathrm{nr}}\, \sqrt{W -
W_{\mathrm{thr}}}\ ,
\end{eqnarray}
where $W_R$ is the resonance mass, $\Gamma_R$ is the full width,
$A_{\mathrm{res}}^2$ and $B_{\mathrm{nr}}$ are the
$Q^2$-dependent magnitudes of the resonant and nonresonant terms,
$K$ is the equivalent real photon energy $[$$K = (W^2 - m^2_p)/(2
m_p)\,$$]$, and ${\mathbf p}^*_{\eta}$ is the three-momentum of
the $\eta$ in the c.m.\ of the $p\,\eta$ system. The $p\,\eta$
production threshold is at $W_{\mathrm{thr}} \approx 1486$~MeV
(in the lowest $W$ bin). At both values of $Q^2$, the fitted
value of the phenomenological nonresonant term
$(B_{\mathrm{nr}}\, \sqrt{W - W_{\mathrm{thr}}})$ was consistent
with zero (with an uncertainty of $1$\% of the resonant term).

The energy-dependent resonance width $\Gamma(W)$ of
Eq.~\ref{eq:breitwig} was parameterized in terms of the branching
fractions $b_{\eta}~(\equiv \Gamma_{\eta}/\Gamma_R$ at $W_R)$,
$b_{\pi}$, and $b_{\pi\pi}$ according to Walker~\cite{Wal69}.  At
present the Particle Data Group (PDG) gives an estimated range
for the $\eta$ branching fraction of $0.30 \le b_{\eta} \le
0.55$~\cite{PDG98}. Therefore, fits to $\sigma_{\mathrm{res}}(W)$
were made assuming three sets of values for the branching
fractions $(b_{\eta}:b_{\pi}:b_{\pi \pi})$, which we define as
Fits~1--3, respectively: $(0.55:0.35:0.10)$, $(0.45:0.45:0.10)$,
and $(0.35:0.55:0.10)$. A consequence of the $p\,\eta$ threshold
is that the fit to $\sigma_{\mathrm{res}}(W)$ cannot constrain
the branching fractions~\cite{Arm98} (\emph{i.e.}, the three fits
result in curves that are virtually indistinguishable).

Based on a branching fraction constraint presented below, we
consider Fit~1 ($b_{\eta} = 0.55$) to $\sigma_{\mathrm{res}}(W)$
to be the preferred fit. The fits for both $Q^2$ points are shown
in Figure~\ref{fig:wdist}. With the Fit~1 branching fractions, we
obtain a full width $\Gamma_R = (154 \pm 20)$~MeV. This width
changed less than 10~MeV over the range of branching fraction
assumptions. The uncertainty is statistical added in quadrature
with systematic. Our result agrees with the PDG estimate
($\approx 150$~MeV)~\cite{PDG98}, and appears lower than the
recent Mainz measurement, $\Gamma_R = (203 \pm
35)$~MeV~\cite{Kru95} (see inset of
Figure~\ref{fig:wdist}). These recent results disagree with the
value of $\Gamma_R = (68 \pm 7)$~MeV obtained from the high-$Q^2$
measurement of Ref.~\cite{Bra84}. The form of the Breit-Wigner
parameterization used by the three groups is essentially the
same, and so does not account for the differences in $\Gamma_R$.

As noted above, the fit to $\sigma_{\mathrm{res}}(W)$ cannot
constrain the branching fraction $b_{\eta}$, but a comparison
between this work and a recent fit to inclusive $(e,e')$
scattering~\cite{Kep94} can. The fit by Keppel \emph{et al.}
models the inclusive cross section in terms of transverse
resonant ($\sigma_{T_{res}}$) and nonresonant ($\sigma_{T_{nr}}$)
contributions using
\begin{equation}
\label{eq:incfit}
\frac{d \sigma}{d\Omega_e dE'_e} = \Gamma_{_T}\, [\,
\sigma_{T_{nr}}\, (1 + \varepsilon\,R_{nr} )\, +\,
\sigma_{T_{res}}\,
]\ .
\end{equation}
In that work, the resonant contribution from each of the three
resonance regions (assumed to be entirely transverse) is fit
using a Breit-Wigner form. The transverse component of the
nonresonant contribution is fit using the phenomenological form
\begin{equation}
\label{eq:transvnonresfit}
\sigma = \sum\limits_{n=1}^3\,C_n(Q^2)\,(W -
W_{\mathrm{thr}})^{n-\frac{1}{2}}\, ,
\end{equation}
where the $C_n(Q^2)$ are fourth-order polynomials in $Q^2$. The
longitudinal component of the nonresonant cross section, which
enters through the longitudinal-to-transverse ratio $R_{nr}$, is
taken from a fit to deep inelastic data~\cite{Whi90}.

The resonant part of the second resonance region is dominated at
low $Q^2$ by the $D_{13}(1520)$. At higher $Q^2$, however, the
$S_{11}(1535)$ begins to dominate, and by $Q^2 = 4$~(GeV$/c)^2$
it is expected that the $S_{11}(1535)$ is responsible for over
$90$\% of the resonant cross section at $W \approx
1535$~MeV~\cite{Bra84}. Assuming that the resonant part of the
inclusive cross section is the incoherent sum of the resonant
contributions of the various decay channels, we can use the
inclusive and exclusive resonant cross sections to put a lower
bound on $b_{\eta}$~\cite{Arm98}:
\begin{equation}
b_{\eta} \ge \frac {\sigma_{\mathrm{res}}(S_{11} \rightarrow
p\,\eta)} {\sigma_{\mathrm{res}}(\mathrm{inclusive})}\ ,
\end{equation}
where both cross sections are taken at $W \approx 1535$~MeV. A
value of $b_{\eta} = 0.55$ results in good agreement between the
high-$Q^2$ point of this work and the inclusive fit; a value of
$b_{\eta} = 0.35$, on the other hand, implies an inclusive cross
section $50$\% \emph{greater} than the fit to the measured
inclusive cross section, which is strong evidence that the
branching fraction is not this low. With the incoherent summation
ansatz given above, and assigning a $10$\% uncertainty to the
inclusive fit, we find a lower bound of $b_{\eta} = 0.45$ with a
$95$\% confidence level. Assuming \emph{complete} $S_{11}$
dominance at $Q^2 = 4.0$~(GeV$/c)^2$, we find a best fit of
$b_{\eta} = 0.55$.

Neglecting resonances other than the $S_{11}(1535)$, we relate
the amplitude $A^p_{1/2}$ to $\sigma_{\mathrm{res}}$
by~\cite{Kru95,PDG76}
\begin{equation}
\label{eq:a12fromsigt}
A^p_{1/2}(Q^2) = \left[\, \frac{W_R\, \Gamma_R}{2\, m_p\,
b_{\eta}}\ \frac {\sigma_{\mathrm{res}}(Q^2,\,W_R)} {1 +
\varepsilon\,R}\, \right]^{1/2} \ .
\end{equation}
Here $\varepsilon$ is the longitudinal polarization of the
virtual photon, and $R = \sigma_{_L} / \sigma_{_T}$. For $R$ we
assumed a parameterization based on a quark-model
calculation~\cite{Rav71}. The expected impact of the
longitudinal-to-transverse ratio $R$ on the final physics result
is small: a $100$\% error in the assumed value $[\approx 4$\% at
$Q^2 = 2.4$~(GeV$/c)^2]$ corresponds to an uncertainty of less
than $1$\% in the quoted value of $A^p_{1/2}\,$.

Table~\ref{tab:finalresults} gives final results for
Fits~1--3. The uncertainties are systematic and statistical added
in quadrature; for $A^p_{1/2}$ we included estimates for the
uncertainties from $\Gamma_R$ and $b_{\eta}$, which were obtained
by varying these quantities over reasonable ranges (150--200~MeV
and 0.45--0.6, respectively) and studying the effect on the
helicity amplitude.


\begin{table}
\caption{Results. The uncertainties are systematic
(\emph{including} estimated uncertainties in $\Gamma_R$ and
$b_{\eta}$ for $A^p_{1/2}$) and statistical added in
quadrature. The top $A^p_{1/2}$ result is for $Q^2 =
2.4$~(GeV$/c)^2$, the bottom is for $Q^2 = 3.6$~(GeV$/c)^2$.
Fit~1 is preferred for reasons discussed in the text. The `best
value' for $b_{\eta}$ assumes $S_{11}$ dominance at $Q^2 =
4$~(GeV$/c)^2$.}
\begin{tabular}{cccc}
 Quantity & Fit 1 & Fit 2 & Fit 3 \\ \tableline
$W_R$ [MeV] \hspace{10pt}  & $1532 \pm 5$ & $1527 \pm 5$ & $1521 \pm 5$ \\
$\Gamma_R$ [MeV] \hspace{10pt} & $154 \pm 20$ & $150 \pm 19$ & $147 \pm 19$ \\
$A^p_{1/2}\ [10^{-3}$~GeV$^{-1/2}]$ &
$50 \pm 7$ & $55 \pm 8$ & $63 \pm 9$ \\
$A^p_{1/2}\ [10^{-3}$~GeV$^{-1/2}]$ &
$35 \pm 5$ & $39 \pm 6$ & $44 \pm 6$ \\
$b_{\eta} = \Gamma_{\eta} / \Gamma_{R}$ &
\multicolumn {3} {c} {$> 0.45$; best value $\approx 0.55$ }\\
\end{tabular}
\label{tab:finalresults}
\end{table}

Table~\ref{tab:uncert} lists the dominant sources of systematic
uncertainty in the measurement and their impact on the
differential cross section and on the helicity amplitude. The
uncertainty in $\frac{d\,^2 \sigma}{d\Omega^*_{\eta}}$ is given
as a range, where the largest uncertainties are for the highest
$W$ bins.

\begin{table}
\caption{Dominant sources of systematic uncertainty, \emph{not}
including $\Gamma_R$ and $b_{\eta}$ (which affect $A^p_{1/2}$).}
\begin{tabular}{lcc}
         & \multicolumn {2} {c} {Fractional uncertainty ($\sigma$) in} \\
Quantity & $\frac{d\,^2 \sigma}{d\Omega^*_{\eta}}$ &
$A^p_{1/2}$ \\ \tableline
Monte Carlo $n\pi$ model & $1$\% to $7$\% & $1$\% \\
$n\pi$ subtraction       & $1$\% to $6$\% & $1$\% \\
Knowledge of $E$         & $1$\% to $10$\% & $0.8$\% \\
Knowledge of $\theta_e$  & $0.2$\% to $11$\% & $1$\% \\
Experimental Acceptance      & $1$\% to $6$\% & $1$\% \\
\end{tabular}
\label{tab:uncert}
\end{table}

Figure~\ref{fig:heltransamp} shows the helicity amplitude
results, along with points calculated from previous $e\,p
\rightarrow e'p\,\eta$ data and some theoretical predictions. All
data points in the figure were calculated using
Eq.~\ref{eq:a12fromsigt} assuming $\Gamma_R = 154$~MeV and
$b_{\eta} = 0.55$; if either assumption is wrong, \emph{all} data
points will scale together. Not included for any of the data
points in the figure are the uncertainties in $\Gamma_R$ and
$b_{\eta}$. Note the good agreement between the high-$Q^2$ point
of the present work and the inclusive fit for $b_{\eta} = 0.55$;
assumption of a lower branching fraction shifts the data
\emph{up} relative to the inclusive fit.

The present result differs from previous work in both the
strength and the slope of the $S_{11}(1535)$ form factor; most
notably, we find a cross section $30$\% lower than that of
Ref.~\cite{Bra84} $[$found by interpolating the results of this
work to $Q^2 = 3$~(GeV$/c)^2$$]$. This difference is reduced in
the amplitude by the square root relating $A^p_{1/2}$ to the
cross section (Eq.~\ref{eq:a12fromsigt}). Although the present
data were taken at a different value of $\varepsilon$ than those
of Ref.~\cite{Bra84}, a longitudinal cross section is not
responsible for the difference between the two measurements; a
value of $R \approx 2.3$ (which is ruled out at low
$Q^2$~\cite{Bre78,Bra78}) would be necessary to account for the
discrepancy.

Of the various CQM curves shown in Figure~\ref{fig:heltransamp},
none exhibit a slope as shallow as that of the data. Those that
indicate an amplitude at $Q^2 \sim 3$~(GeV$/c)^2$ roughly
consistent with experimental data also predict excess amplitude
at lower $Q^2$. Our data also have consequences for a recent
coupled-channel model for the $S_{11}(1535)$~\cite{Kai95}; the
proposed quasi-bound $K\Sigma$ (five quark) state is expected to
have a form factor that decreases more rapidly than is observed.

Figure~\ref{fig:q3a12} shows the quantity $Q^3 A^p_{1/2}$ for the
$S_{11}(1535)$, which is predicted by pQCD to asymptotically
approach a constant at high $Q^2$~\cite{Car88}. As has been
pointed out elsewhere~\cite{ISG84}, such scaling might be due to
non-perturbative contributions. While there is no \emph{strong}
scaling evident in the figure, our data indicate that $Q^3
A^p_{1/2}$ may be approaching a constant value by $Q^2 \sim
5$~(GeV$/c)^2$.

\section{Conclusions}
\label{sec:conc}

We have presented the results of a precise, high statistics
measurement of the $e\,p \rightarrow e'p\,\eta$ process at $W
\approx 1535$~MeV and at $Q^2 = 2.4$ and
3.6~(GeV$/c)^2$~\cite{angdisdat}.  The contribution of terms
other than $S$-wave multipoles is observed to be less than $7$\%,
which is consistent with previous measurements.  More
importantly, the cross section obtained from the new data is
about $30$\% lower and indicates a full width twice that of the
only other exclusive measurement at comparable
$Q^2$~\cite{Bra84}.

While the new data exhibit no strong perturbative signature, they
do have a $Q^2$ dependence that is markedly different than the
older high-$Q^2$ measurement. Even given the new (lower) cross
section obtained from this measurement, however, relativized
versions of the quark model fail to reproduce the $Q^2$
dependence seen experimentally.

A comparison of the new high-$Q^2$ datum (the highest in
existence) with a recent inclusive analysis indicates an
$S_{11}(1535) \rightarrow p\,\eta$ branching fraction of at least
$b_{\eta} =0.45\,$. Using $b_{\eta} = 0.55$ we obtain $\Gamma_R =
154$~MeV and a new measurement of $A^p_{1/2}(Q^2)$ (see
Table~\ref{tab:finalresults}).


We wish to acknowledge the support of those in the Jefferson Lab
accelerator division for their invaluable work during the
experiment. This work was supported in part by the U.\ S.\
Department of Energy and the National Science Foundation. CSA
also thanks SURA and Jefferson Lab for their support.


\begin{figure}
\caption{A plan view of the Hall~C end station at Jefferson Lab.
The electron beam enters from the left, and the scattering takes
place in the cryogenic target placed in the beamline.  In this
experiment, outgoing particles were detected by two magnetic
spectrometers: the Short-Orbit Spectrometer (SOS) was used to
detect electrons and the High-Momentum Spectrometer (HMS) was
used to detect protons.
\label{fig:hallcfloorplan}}
\end{figure}

\begin{figure}
\caption{A side view of the HMS detector stack, as seen from the
door of the detector hut. The detected particles travel from
left to right (along positive $z$).
\label{fig:hdetstack}}
\end{figure}

\begin{figure}
\caption{The response of the SOS calorimeter and \v{C}er\-en\-kov
for events of a typical data run. The calorimeter response
$E_{\mathrm{cal}}$ is the total energy deposited normalized to
the particle momentum, while the \v{C}er\-en\-kov response
$N_{\mathrm{p.e.}}$ is the number of photo-electrons
detected. The events at $N_{\mathrm{p.e.}}  = 0$ are $\pi^-$
(note the peak at $E_{\mathrm{cal}} \approx 0.25$). The events at
$N_{\mathrm{p.e.}} > 0$, $E_{\mathrm{cal}} > 0.7$ are electrons.
The events at $N_{\mathrm{p.e.}} > 0$, $E_{\mathrm{cal}} \approx
0.3$ are caused by $\pi^-$ that produced knock-on electrons that
triggered the \v{C}er\-en\-kov. Note that the $z$ axis is on a
log scale.
\label{fig:sospid2d}}
\end{figure}

\begin{figure}
\caption{Velocity from time of flight ($\beta_{\mathrm{HMS}}$)
and coincidence time (the difference in time of arrival for the
two spectrometers) for events of a typical data run. The band of
events at $\beta_{\mathrm{HMS}} \approx 1$ are $\pi^+$, while
those at $\beta_{\mathrm{HMS}} \approx 0.8$ are protons. The real
proton coincidences are at $t = 0$~ns, and the nominal 2~ns radio
frequency structure of the beam is visible in the adjacent
accidental peaks. The low-$\beta_{\mathrm{HMS}}$ tail emanating
from the real coincidence peak is most likely due to protons
undergoing interactions in the detectors after the drift
chambers.
\label{fig:cointime2d}}
\end{figure}

\begin{figure}
\caption{A plot of $M_{x}^2$ for one kinematic setting. The peak
at $M_{x}^2 \approx 0.3$~(GeV$/c^2)^2$ corresponds to undetected
$\eta$ mesons in the final state (the peak at $M_{x}^2 \approx
0.02$~(GeV$/c^2)^2$ corresponds to $\pi^0$, the subject of
Ref.~\protect \cite{Fro98}).  Note the presence of the multi-pion
background as well as the radiative tail extending to the right
of the $\eta$ peak.
\label{fig:fridgeplot}}
\end{figure}

\begin{figure}
\caption{Fits to the $M_{x}^2$ distribution for several typical
($\cos{\theta_{\eta}^*}\,$, $\phi^*_{\eta}$) bins, one kinematic
setting.
\label{fig:indivfit}}
\end{figure}

\begin{figure}
\caption{Results of background fits for several typical kinematic
settings. Data are on the left and the corresponding Monte Carlo
result is on the right.  Each figure shows the integration of
sixty individual ($\cos{\theta_{\eta}^*}\,$, $\phi^*_{\eta}$)
bins and their respective fits (like those shown in
Figure~\ref{fig:indivfit}). The solid line is the sum of the
background and peak fits; the dashed line shows the background
only. The lines at the bottom of the data plots show the small
contribution from the accidental coincidence and HMS collimator
backgrounds.
\label{fig:bgfit1}}
\end{figure}

\begin{figure}
\caption{Angular distributions for the $Q^2 = 3.6$~(GeV$/c)^2$
data. Each plot shows the $\cos{\theta^*_{\eta}}$ distribution
for a single ($W$, $\phi^*_{\eta}$) bin. The rows correspond to
different bins in $W$, the columns to different bins in
$\phi^*_{\eta}\,$. Data corresponding to $\phi^*_{\eta} = \pm
90$~degrees are not shown; the out-of-plane experimental coverage
was complete only for the lowest $W$ bin (where the data looked
similar to that in the $\phi^*_{\eta}$ bins shown here), and was
almost nonexistent at higher $W$. The lines are $S$-wave fits to
the data.
\label{fig:angdist5pass}}
\end{figure}

\begin{figure}
\caption{Fit~1 to $\sigma_{\mathrm{res}}(W)$ for the two $Q^2$
points of this work (errors on the data are statistical
only). Note the presence of the $p\,\eta$ threshold. The inset
shows the $W$-dependence of this cross section as measured by the
present work (solid line, $\Gamma_R = 154$~MeV), Ref.~\protect
\cite{Kru95} (dashed line, $\Gamma_R = 203$~MeV), and
Ref.~\protect \cite{Bra84} (dotted line, $\Gamma_R =
68$~MeV). The curves in the inset have been normalized to the
same magnitude.
\label{fig:wdist}}
\end{figure}

\begin{figure}
\caption{The helicity amplitude $A^p_{1/2}(Q^2)$ of the
$S_{11}(1535)$, measured via $e\,p \rightarrow e'p\,\eta$,
together with some theoretical predictions.  The data points
(\protect\cite{Kru95,Bra84,Bre78,Bra78,Kum73,Bec74,Ald75} and the
present work) were calculated using $\Gamma_R = 154$~MeV,
$b_{\eta} = 0.55$, and the parameterization of $R$ referenced in
the text. The errors shown on previous data are statistical
only. The errors shown for the present work include both
statistical \emph{and} systematic uncertainties, with the
exceptions noted in the text. The theoretical curves of
Refs.~\protect \cite{Cap98,Sta95,Li90,War90,Aie98} are based on
variants of the CQM. The curve from Ref.~\protect \cite{Car88} is
the result of a pQCD calculation. The curve from Ref.~\protect
\cite{Kep94} is a fit to inclusive data.
\label{fig:heltransamp}}
\end{figure}

\begin{figure}
\caption{The quantity $Q^3 A_{1/2}^p(Q^2)$ for the
$S_{11}(1535)$. The dot-dashed line is an exponential fit to the
cross section given by the two points of the present work
($\sigma_{\mathrm{res}} = 16.5\, \exp{[-0.565\, Q^2]}~\mu$b,
where $Q^2$ is in $[$(GeV$/c)^2]$), and the solid line is a fit
to inclusive data (as in Figure~\protect
\ref{fig:heltransamp}). The errors that are shown, and the
assumed values for $\Gamma_R\,$, $b_{\eta}\,$, and $R$, are the
same as in Figure~\protect \ref{fig:heltransamp}.
\label{fig:q3a12}}
\end{figure}

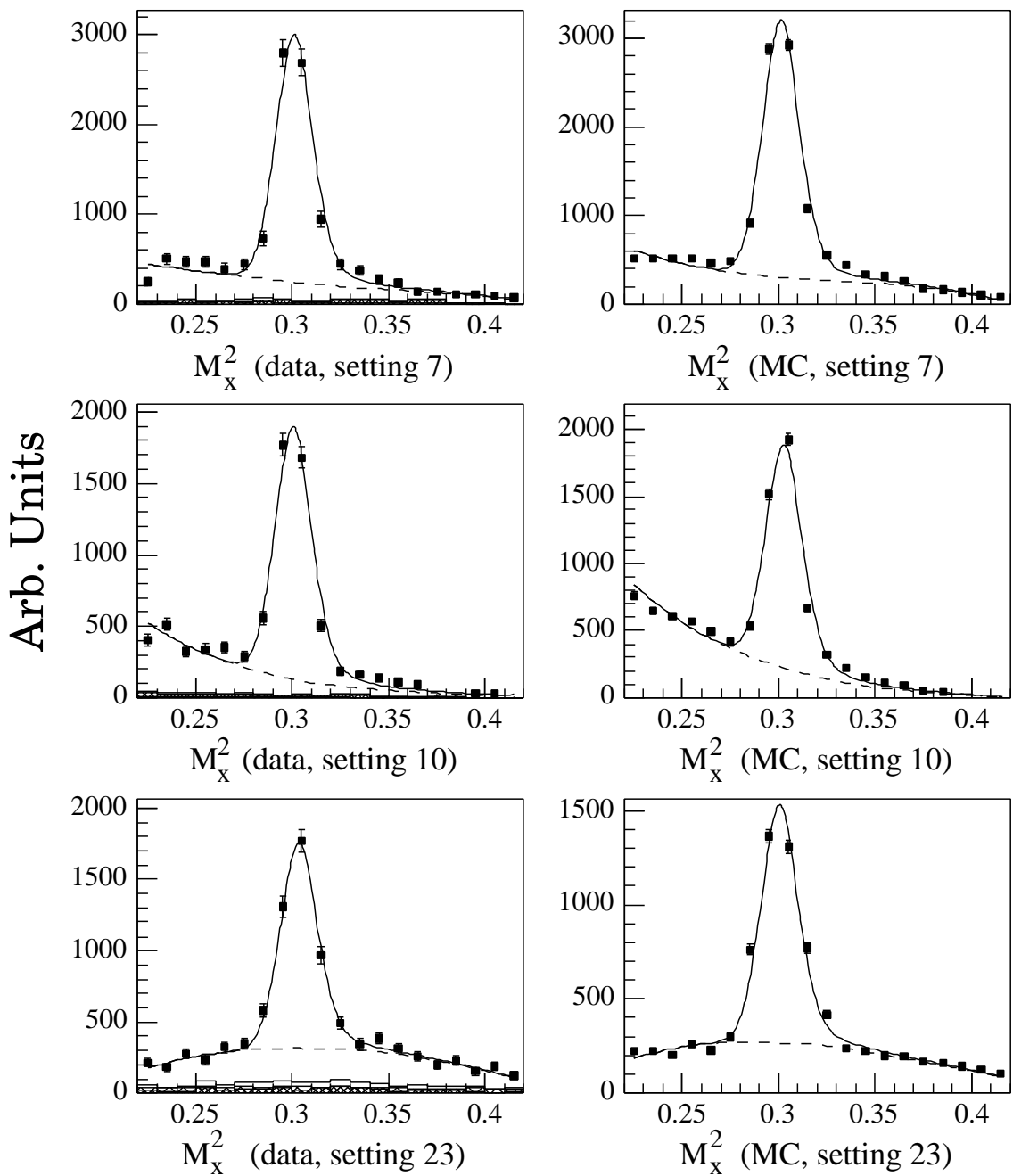
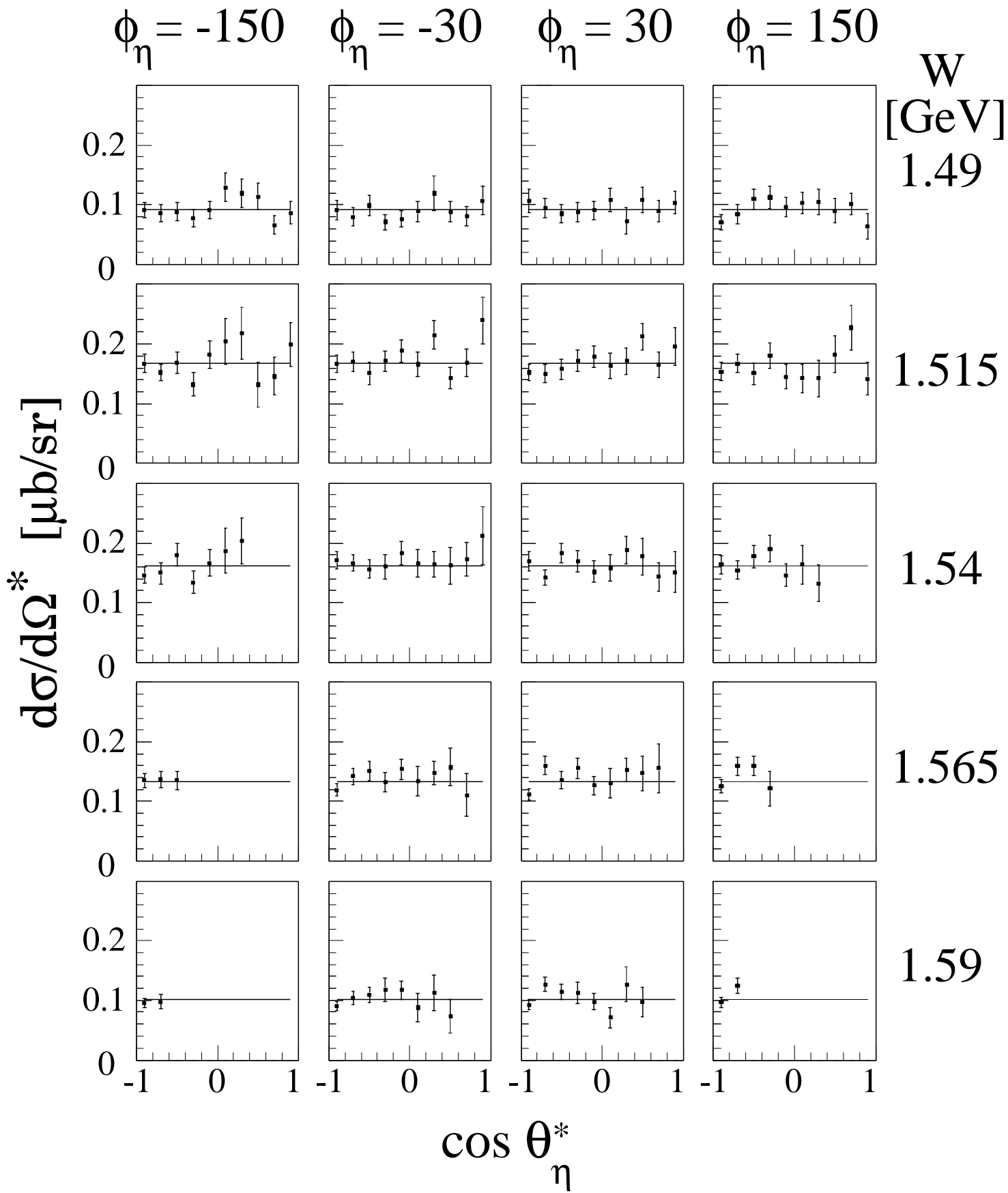
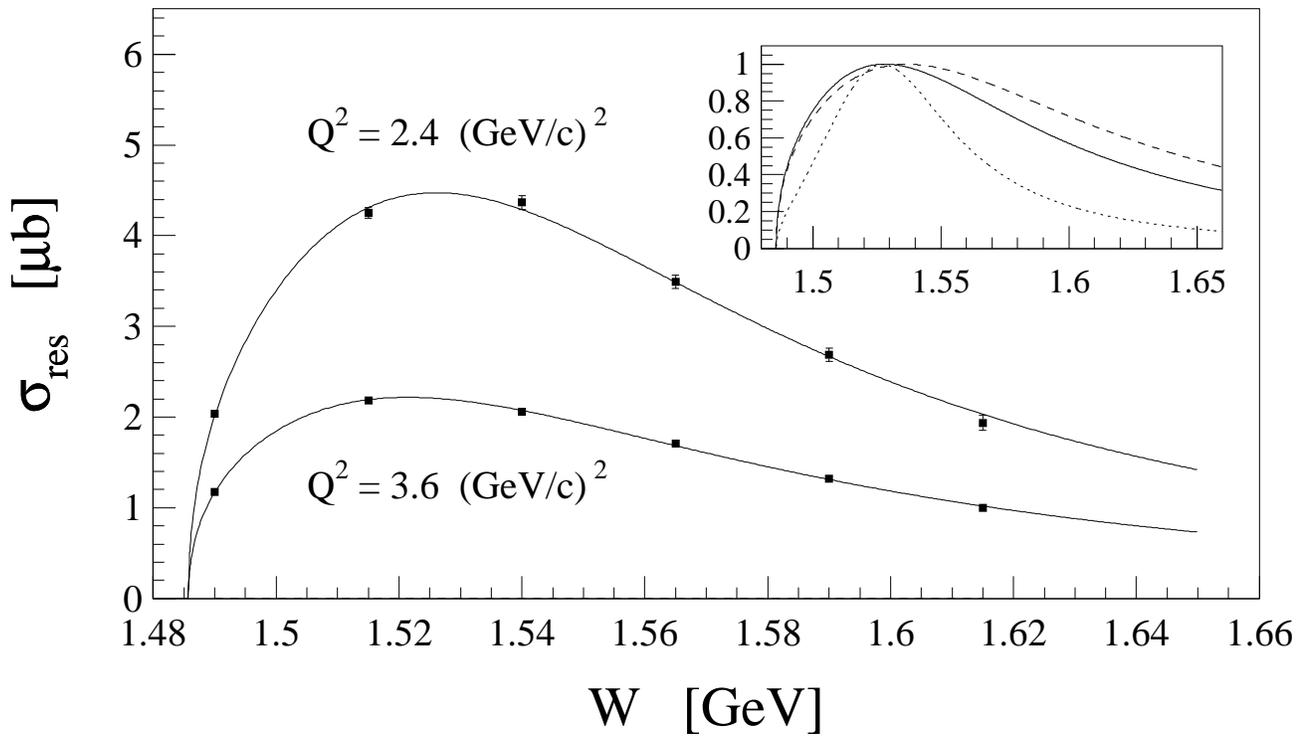
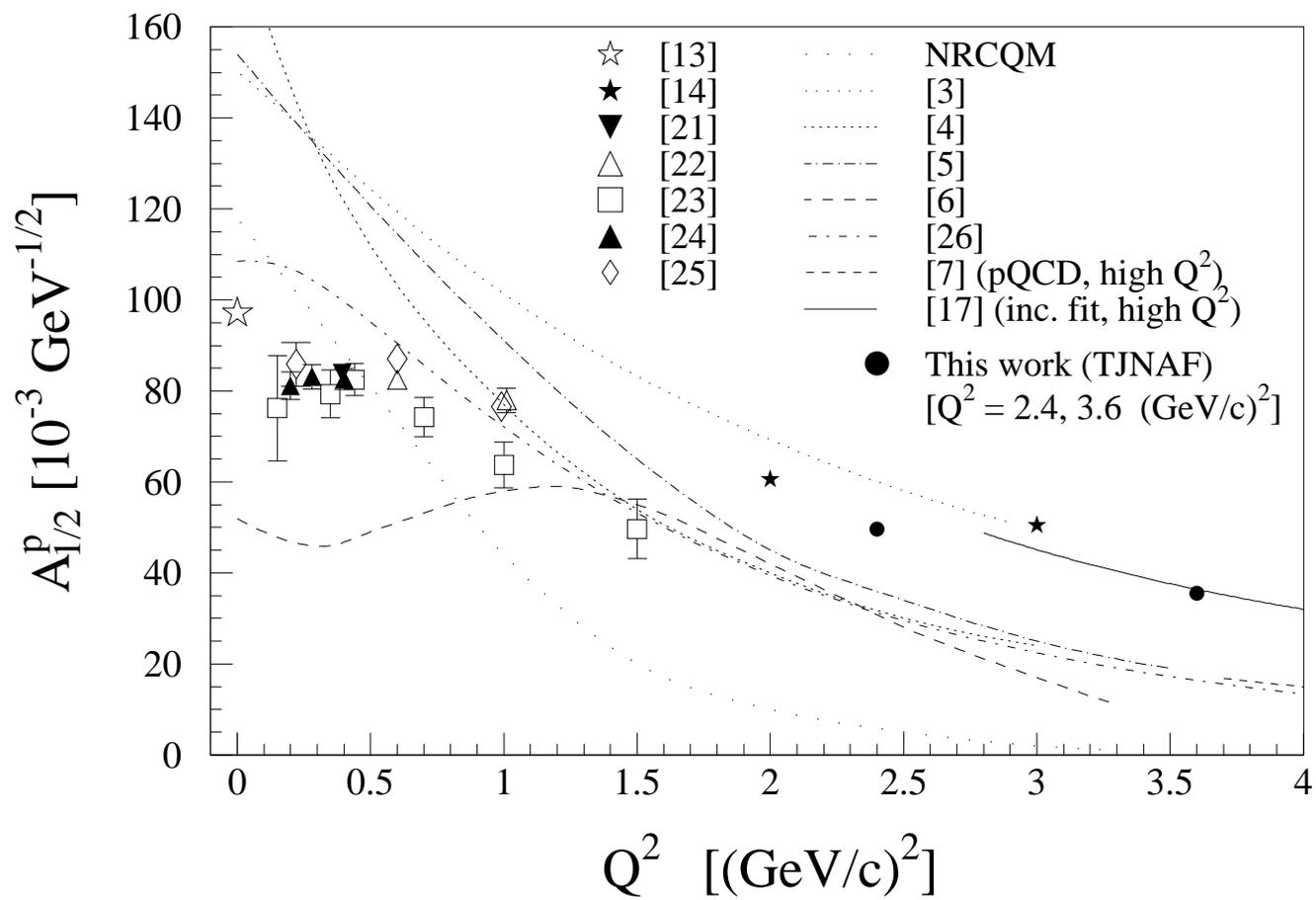
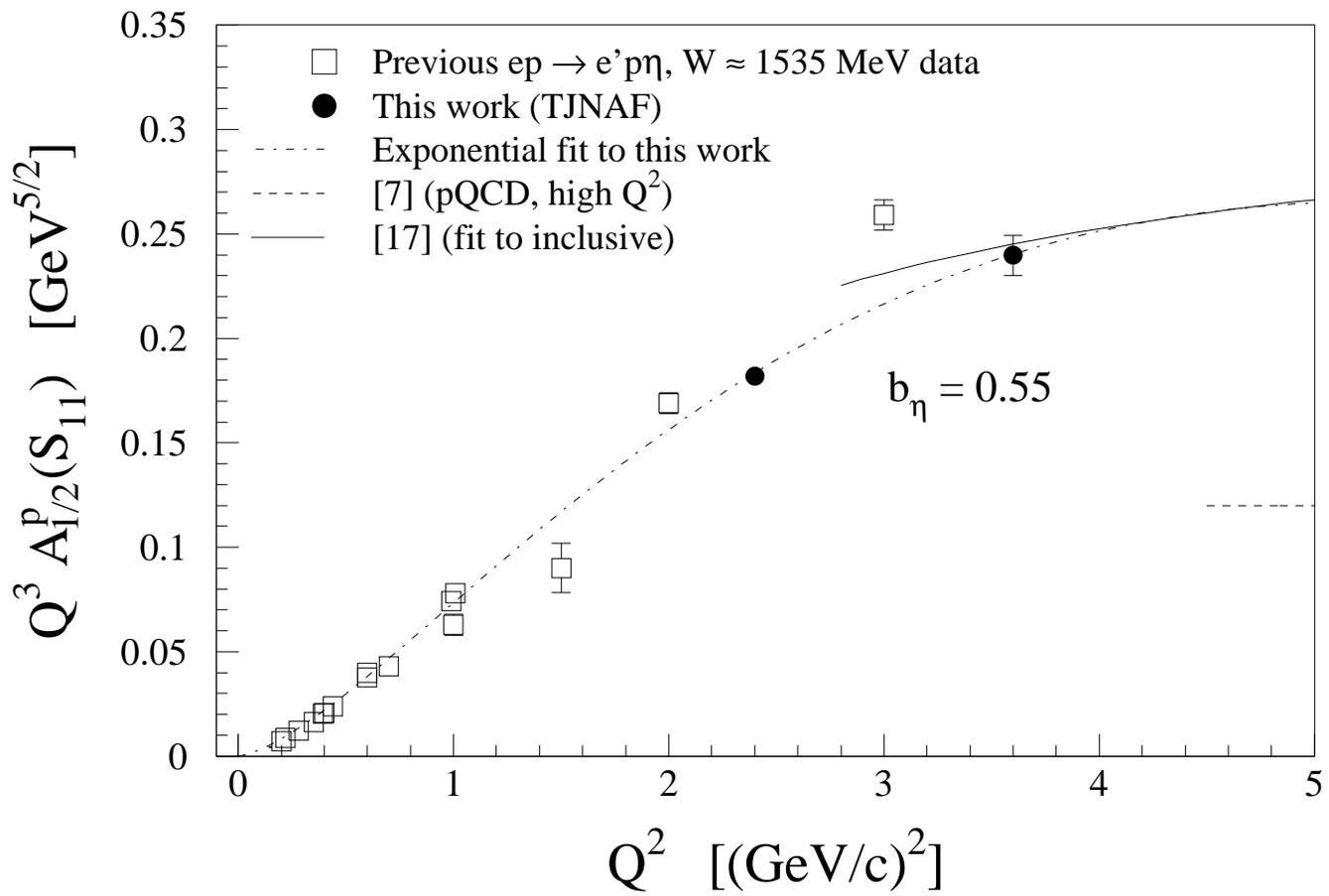
\end{document}